\documentclass[aps, prd, preprint, nofootinbib, amsfonts, floatfix]{revtex4}

\usepackage{graphicx}
\usepackage{amsmath,amsfonts}
\usepackage{epsfig,color}
\usepackage[dvipsnames]{xcolor}

\def\ttl#1{{\it #1}}

\newcommand{\ra}{\:\rightarrow\:}
\newcommand{\Lar}{\mathcal{L}}
\newcommand{\SU}{\text{SU}}
\newcommand{\BigO}{\mathcal{O}}
\newcommand{\dd}{\text{d}}
\newcommand{\be}{\begin{eqnarray}}
\newcommand{\ee}{\end{eqnarray}}

\newcommand\nn{\nonumber}
\newcommand{\Tr}{{\rm Tr}}

\newcommand{\mat}{\left ( \begin{array}{cc}}
\newcommand{\emat}{\end{array} \right )}
\newcommand{\vect}{\left ( \begin{array}{c}}
\newcommand{\evect}{\end{array} \right )}

\definecolor{red}{rgb}{1.00, 0.00, 0.00}

\definecolor{blue}{rgb}{0.00, 0.00, 1.00}
\definecolor{green}{rgb}{0.20, 0.6, 0.1}

\definecolor{darkgreen}{rgb}{0.0, 0.4, 0.0}

\let\ifcomments\iftrue
\def\commentsoff{\global\let\ifcomments\iffalse}
\let\commentsize\small
\def\tinycomments{\global\let\commentsize\footnotesize}

\long \def \blockcomment #1\endcomment{}

\def\dchpt{\lowercase{d}$\chi$PT}
\def\chpt{$\chi$PT}
\def\mathdchpt{{\rm d\chi PT}}
\def\mathchpt{{\rm \chi PT}}
\def\Eq#1{Eq.~(\ref{#1})}
\def\Eqs#1{Eqs.~(\ref{#1})}
\def\det{{\rm det\,}}

\begin{document}

\title{The \begin{boldmath}$\epsilon$-regime\end{boldmath} of dilaton chiral perturbation theory}

\author{Taro V. Brown,${}^1$ Maarten Golterman,${}^2$ Svend Kr{\o}jer,${}^1$ Yigal Shamir,${}^3$ K. Splittorff${}^1$}
\affiliation{${}^1$The Niels Bohr Institute, University of Copenhagen, Blegdamsvej 17, DK-2100, Copenhagen {\O}, Denmark \\ ${}^2$Department of Physics and Astronomy, San Francisco State University, San Francisco, CA 94132, USA \\
${}^3$Raymond and Beverly Sackler School of Physics and Astronomy, Tel Aviv University, 69978 Tel Aviv, Israel}

\date{\today}
\begin{abstract}
The $\epsilon$-regime of dilaton chiral perturbation theory is introduced.
We compute the dilaton mass, the
chiral condensate and the topological susceptibility in the $\epsilon$-regime,
as a function of the fermion mass. The microscopic spectral density
of the Dirac operator is obtained from dilaton chiral perturbation theory. 
Our main result is that the chiral condensate and the spectral density
are related to their counterparts from ordinary chiral perturbation theory
via a simple scaling relation.  This relation originates from
the mass dependence of the dilaton potential, and is valid 
in both the $\epsilon$-regime and the $p$-regime. 
In the $\epsilon$-regime, moreover, all 
results agree with the universal predictions to leading order in $\epsilon$.
\end{abstract}

\maketitle

\section{Introduction}

Lately, a large number of studies have appeared of QCD-like theories using
lattice techniques originally developed for studying QCD,
for reviews see Refs.~\cite{DeGrand:2015zxa,Nogradi:2016qek,Pica:2017gcb,Svetitsky:2017xqk}.
In particular, theories where the number of fermions or their representation
suggest that the theory could be in a (confining but) near-conformal phase
have drawn a lot of attention. One of the interesting outcomes
is the observation of a light scalar particle in addition to the usual mesonic
modes (the ``pions''), see Refs.~\cite{Aoki:2016wnc,Appelquist:2016viq,Appelquist:2018yqe,Fodor:2017nlp,Fodor:2019vmw}
for recent examples. The possible nature of this additional mode is still
an open question; an exciting possibility is that it is a pseudo
Nambu--Goldstone mode associated with approximate dilatation symmetry
(see the above reviews). A fruitful way to address this question
is to include the additional mode within the framework of
low energy effective field theory (EFT), and match the predictions thereof
with the results obtained from the underlying lattice theory.
Possible low-energy effective theories which take into account
both chiral and conformal symmetry go back to Refs.~\cite{Ellis,ISS}.
Recently, in Refs.~\cite{GS,GSlat,GStr,GSlarge-m}, a systematic
low energy theory was formulated as an extension of chiral perturbation theory
assuming that the additional light scalar mode is a pseudo Nambu--Goldstone
mode associated with the almost restored dilatation invariance.
The first tests of this effective theory against lattice results are
encouraging \cite{Appelquist:2017wcg,Appelquist:2017vyy,Fodor:2019vmw,GSlarge-m,GSlat19},
and we will follow this approach in the present work.

The $\epsilon$-regime of chiral perturbation theory (\chpt) \cite{GLthermo,LS}
has lead to a surprising wealth of insights into QCD, and the interplay between
numerical lattice QCD and analytic results from the $\epsilon$-regime of \chpt\
has helped developing both lattice QCD and chiral perturbation theory.
For example the analytic analysis of order $a^2$ effects
\cite{SharpeSingleton,BNS,RS,BRS,DSV,KSV,S-plen-lat,Shindler:2009ri} has explained why
the width of the distribution of the smallest eigenvalues of the
(massive) hermitian Wilson--Dirac operator scales with the square root
of the volume close to the continuum, as first observed
on the lattice \cite{Luscher}.
Likewise the analytic analysis for non-zero quark chemical potential
\cite{AOSV,OSV,SV-phase,SSvet} in the $\epsilon$-regime has given substantial
new insights into the sign problem in lattice QCD.

In this paper we define and perform a first study of the $\epsilon$-regime
of chiral perturbation theory extended to include the dilatonic meson (\dchpt).
One technical advantage of the $\epsilon$-regime of \dchpt\ as compared to
the $p$-regime studies in Refs.~\cite{GS,GSlat,GSlarge-m} is
(as in ordinary \chpt\ \cite{GLthermo,LS,Toublan:2000dn,Factorization})
the possibility of evaluating the partition function explicitly.
In order to extend this property to \dchpt\ we define the new power counting
such that effectively the pions are in the $\epsilon$-regime,
while the dilaton is kept in the $p$-regime. As discussed below,
other counting schemes are possible, but, because the dilaton is described by
a non-compact field, they lead to integrals over the dilaton potential 
that include asymptotic field values which are outside the domain of
our effective field theory.
Our chosen counting allows us to address the influence of the dilatonic meson
on the fermion mass dependence of the topological susceptibility,
the chiral condensate and the average Dirac spectrum.

At leading order in the $\epsilon$-counting the results all agree with
the universal predictions of ordinary chiral perturbation theory.
The new dependence induced by the dilaton enters at next to leading order,
when the pion mass is in the $\epsilon$-regime.
We find that the chiral condensate and the spectral density
are related to their counterparts from ordinary chiral perturbation theory
via a simple scaling relation.  This relation can be traced back to
the mass dependence of the dilaton potential, and, as it turns out,  is valid 
both the $\epsilon$-regime and the $p$-regime.

The paper is organized as follows. After a brief review of
dilaton chiral perturbation theory in Sec.~\ref{sec:dCPT} we discuss
the possible counting schemes in Sec.~\ref{sec:counting} and reanalyze
the structure of the dilaton potential in Sec.~\ref{sec:minimum}.
Section \ref{sec:epsilon} contains the results obtained in
the $\epsilon$-regime of \dchpt~for respectively the dilaton mass,
the topological susceptibility and the chiral condensate.
In Sec.~\ref{sec:Dirac} we derive the average eigenvalue density for
the massless Dirac operator. Finally, Sec.~\ref{sec:conc} sums up the conclusions.
Some technical details regarding the calculation of
the topological susceptibility are relegated to App.~\ref{topodetail},
while a few explicit results for the partition function and the
generating functional are derived in App.~\ref{largeN}.

\section{Dilaton chiral perturbation theory}
\label{sec:dCPT}

In addition to the pseudo-scalar Nambu--Goldstone bosons associated with chiral symmetry
breaking, which we will refer to as
``pion'' fields from standard \chpt, dilaton chiral perturbation theory includes a real scalar field $\tau$ representing the light dilatonic meson (``dilaton,'' for short) associated with weak breaking of dilatation invariance.  This effective field theory was derived in Ref.~\cite{GS} on the basis of the
assumption that the dilaton can be interpreted as a pseudo Nambu--Goldstone boson associated
with a weak breaking of dilatation symmetry, postulated to occur outside, but in the vicinity of the
conformal window in theories with $N_f$ fermion flavors in the fundamental representation of the
gauge group.   The connection between the underlying gauge theory and the effective theory is
made through the introduction of sources $\chi$ and $\sigma$ for the pions and dilaton, respectively.
In this construction, the sources are endowed with transformations under the relevant symmetry
group, elevating them to spurions.
Here we will review the transformation properties of  the pion and dilaton fields as well as the  spurion fields $\chi(x)$ and $\sigma(x)$.
 The possible terms in the EFT, \dchpt, are determined by the invariance properties under these transformations.

We start off with the chiral transformations. For the pion field $U(x)\in\SU(N_f)$
 we have the standard $\SU(N_f)\times\SU(N_f)$ transformation \cite{GL}
\be \label{eq:Utrans}
U\ra g_{L} U g_{R}^\dagger\ , \qquad g_{L}\in \SU(N_f)_{L}\ ,\quad g_{R}\in \SU(N_f)_{R}\ ,
\ee
while the $\tau$ field is unaffected.

The fermion mass explicitly breaks this symmetry and to encode this into the effective theory
it is promoted to the associated spurion $\chi$, which transforms as
\be \label{eq:mtrans}
\chi\ra g_{L} \chi g_{R}^\dagger
\ .
\ee
After the construction of the EFT, the spurion field $\chi$ is set equal to the fermion mass $m$.

Under dilatations, the spurion $\sigma$ transforms as \cite{GS}
\be
\label{eq:sigmatrans}
\sigma(x)\ra \sigma(\lambda x)+\log\lambda\ ,
\ee
where $\lambda$ is the scale factor.   Just like the pion field $U$ transforms like the source $\chi$,
the effective dilaton field $\tau$ transforms like $\sigma$, and, in addition to
Eq.~(\ref{eq:sigmatrans}) dilatation transformations act on all
fields as
\begin{eqnarray}
\label{eq:diltrans}
    U(x)&\ra&U(\lambda x)\ ,\\
    \chi(x)&\ra&\lambda^{4-y}\chi(\lambda x)\ ,\nonumber\\
    \tau(x)&\ra&\tau(\lambda x)+\log \lambda\ , \nonumber
\end{eqnarray}
where $y=3-\gamma_*$ with $\gamma_*$ the mass anomalous dimension at the nearby
infrared fixed point \cite{GS}.
We note that it is the renormalized mass spurion $\chi(x)$ which transforms this way \cite{GStr}.
It will be sufficient for our purposes
to consider only a constant dilatation spurion field, $\sigma(x)=\sigma_0$, which transforms as $\sigma_0\ra\sigma_0+\log\lambda$.

We will consider the theory in a volume ${\cal V}=L^4$ with periodic boundary
conditions, which also breaks dilatation symmetry explicitly.  However, again the symmetry can be restored formally if we let also $L$,  the side-length of the box, be a spurion transforming as
\be
L \to \frac{L}{\lambda} \ .
\label{L-spurion-transf}
\ee

Using these fields the simplest non-trivial Lagrangian $\Lar$ for which the action $S=\int d^4x\, \Lar(x)$ is invariant under the stated transformations of fields and spurions is \cite{GS}
\be
\Lar=\Lar_\pi+\Lar_\tau+\Lar_m+\Lar_d\ ,
\label{Leff}
\ee
with (working in Euclidean space)
\be
\label{eq:lagrangian}
\Lar_\pi &=& \frac{f^2_\pi}{4}\,V_\pi(\tau-\sigma_0)\,e^{2\tau}\Tr[\partial_\mu U^\dagger\partial^\mu U]\ , \\
\Lar_\tau &=& \frac{f^2_\tau}{2}\, V_\tau(\tau-\sigma_0)\,e^{2\tau}\partial_\mu \tau \partial^\mu \tau\ ,\\
\Lar_m &=& -\frac{f^2_\pi B_\pi}{2}\, V_m(\tau-\sigma_0)\,e^{y\tau}\Tr[\chi^\dagger U+U^\dagger \chi]\ ,\\
\Lar_d &=&f^2_\tau B_\tau\, V_d(\tau-\sigma_0)\,e^{4\tau} \ ,
\ee
where $V_i$, $i=\pi,\tau,m,d$ are invariant potentials.
The low energy constants (LECs), $f_\pi$, $f_\tau$, $B_\pi$ and $B_\tau$,  are invariant under both chiral and dilatation transformations. 
The finite-volume action is invariant, too,
provided that the linear size of the box spurion transforms as in Eq.~(\ref{L-spurion-transf}).

The functional forms of $V_\pi$, $V_\tau$, $V_m$ and $V_d$ are unconstrained by symmetry and call for a counting scheme to be introduced.  We will turn to this in the next section.

\begin{boldmath}
\section{$\epsilon$-counting and leading order Lagrangian}
\label{sec:counting}
\end{boldmath}

In this section we will introduce the $\epsilon$-counting scheme and identify the lowest order Lagrangian.  As we will hold fixed the number of flavors, $N_f$, {\it i.e.}, the ``distance'' to the
conformal window, the amount by which dilatation symmetry is broken (in the massless theory) is also fixed.
In current simulations \cite{Aoki:2016wnc,Appelquist:2016viq,Appelquist:2018yqe,Fodor:2017nlp,Fodor:2019vmw}, the dilaton mass is in the $p$-regime.
We will thus assume that the dilaton is in the $p$-regime, but consider the situation in which
the fermion mass $m$ is chosen small enough that the pion mass is in the $\epsilon$-regime. As we will explain in section \ref{sec:Dirac}, this counting is also natural when we consider the
valence and ghost quarks of the generating functional for the Dirac eigenvalues.

\begin{table}[t]
\centering
\begin{tabular}{c|c|c}
\hline
Term & \ $p$-counting\; & \ $\epsilon$-counting \\
\hline\hline
$\partial_\mu,\frac{1}{L}$ & $\BigO(p)$  & $\BigO(\epsilon)$\\
\hline
$m$ & $\BigO(p^2)$ & $\BigO(\epsilon^4)$\\
\hline
$n_f-n_f^*$ & $\BigO(p^2)$ & $\BigO(\epsilon^2)$ \\
\hline
\end{tabular}
\caption{The $p$-counting of Ref.~\cite{GS} and the $\epsilon$-counting introduced here.}
\label{tab:counting}
\vspace*{3ex}
\end{table}

\vspace{2ex}
\begin{boldmath}
\subsection{Definition of the $\epsilon$-counting}
\label{sec:def-counting}
\end{boldmath}

In Ref.~\cite{GS} the $p$-regime was considered. The small parameters in this counting are
\be
m\sim n_f-n_f^*\sim p^2\sim1/L^2\sim\delta\ll 1\ , &&
\qquad (\mbox{$p$-regime})
\\
1/N_f\sim1/N_c\equiv 1/N\ll 1 \ , &&
\label{Vlim}
\ee
with
\be
n_f=\lim_{N_f, N_c\to \infty}\frac{N_f}{N_c}\ ,\qquad n_f^*=\lim_{N_c\to \infty}\frac{N_f^*(N_c)}{N_c} \ ,
\label{eq:ep-counting}
\ee
where $N_f^*(N_c)$ is the largest number of fundamental-representation
flavors for which the $\SU(N_c)$ gauge theory still confines.
The difference $n_f-n_f^*<0$, which measures the distance from the conformal window,
controls the explicit (hard) breaking of scale invariance coming from
the running of the gauge coupling.  This is similar to the way $m$ controls
the explicit breaking of chiral symmetry in standard \chpt\ \cite{GL}.
(Of course, $m$ breaks scale invariance as well.)
Throughout this paper we invoke the Veneziano limit~(\ref{eq:ep-counting})
in order to treat $n_f-n_f^*$ as a continuous parameter.
This requires $1/N \le |n_f-n_f^*| \sim \delta$, but otherwise
we will not demand any particular relation between $1/N$ and $\delta$.

In this paper we instead consider \dchpt\ in a different counting, in which the pion mass
is in the $\epsilon$-regime. There are different possible choices; we choose to define the $\epsilon$-counting
\be \label{ep-delta-counting}
m\sim p^4\sim1/L^4\sim\epsilon^4\ll 1\ ,  &&
\qquad (\mbox{$\epsilon$-regime}) \ , \\
n_f-n_f^*\sim\delta\ll 1\ . &&  \nn
\ee
This corresponds loosely to treating the pion field in $\epsilon$-counting while the dilaton field is kept in $p$-counting.  For small $m$, we have $m_\pi^2\sim m$, and so this choice implies $m_\pi L\ll 1$.
Likewise, $M_\tau^2\sim |n_f-n_f^*|$, and choosing
\be
\delta\sim\epsilon^2\sim p^2\ ,
\ee
keeps the dilaton in the $p$-regime.

A comparison of the $p$- and $\epsilon$-countings is shown in table \ref{tab:counting}.
LECs are always order 1.  As for the Veneziano limit, we have that
$f_\pi\sim \sqrt{N}$, $f_\tau\sim N$, while $B_\pi$ and $B_\tau$ are $\BigO(1)$,
ensuring that the pion and dilaton masses are $\BigO(1)$ in the large-$N$
counting as well.

When using $\epsilon$-counting in standard \chpt\ the partition function separates into a group integral over the zero-modes and a path integral over the non-zero modes \cite{GLthermo}.
As we will show below this still holds with the added dilatonic terms, at least when the dilaton
is kept in the $p$-regime.   Table~\ref{tab:progressindilatonic} shows the various choices one can
make for the counting of both pion and dilaton degrees of freedom, along with the choice
we make in this paper, in which $\delta\sim\epsilon^2$.

In principle, one might also consider
a theory with the dilaton in the $\epsilon$-regime, by choosing $\delta\sim\epsilon^4$.
However, in this case, it is not clear how to proceed.   An additional
assumption about the behavior of $V_d$ for $\tau\to-\infty$ would be needed, and higher order terms would have to come into play, because, with the form of
$V_d$ discussed below in Sec.~\ref{sec:minimum}, the integral over $\tau$
diverges in that region.   This is beyond the scope of the present paper, in which we
keep the dilaton in the $p$-regime, and thus we only need to consider
small fluctuations of the $\tau$ field, including for its zero mode.
This assumption is only self-consistent in the $p$-regime.

\begin{table}[t]
\centering
\begin{tabular}{c||c|c}
\hline
$\pi${\large\textbackslash}$\tau$
& $p$ & $\epsilon$  \\
\hline\hline
$p$  & \cite{GS,GSlat} & \ Full $V_d$-dependence  \\
\hline
$\epsilon$ & \ This paper\; & \ Full $V_d$-dependence \\
\hline
\end{tabular}
\caption{The 4 possible combinations of $p$- and $\epsilon$- counting for the $\pi$ and $\tau$ fields. The two combinations on the right hand side are sensitive to the form of $V_d$ and have yet to be explored.}
\label{tab:progressindilatonic}
\vspace*{3ex}
\end{table}

\vspace{2ex}
\section{The minimum of the dilaton potential in the chiral limit}
\label{sec:minimum}

In the $\epsilon$-regime of standard \chpt\ it is advantageous to express the pion field through its zero mode part $U_0$ and its non-zero mode part $\xi(x)$ \cite{GLthermo}. The $U_0$ part of the standard-\chpt\ path integral factorizes and may then be performed independently to obtain the functional dependence of the partition function on the quark mass, to leading order in $\epsilon$.
In order to see how this works in the presence of the dilaton in \dchpt, we first need to consider
the potentials $V_i$ appearing in the Lagrangian~(\ref{Leff}), in particular, $V_m$ and $V_d$.
 In the following we discuss the minimum of the dilaton potential in the chiral limit; we will return to the effective dilaton potential for non-zero quark mass in Sec.~\ref{sec:epsilon}.
We will use the counting scheme defined in Eq.~(\ref{ep-delta-counting}).

It was shown in Ref.~\cite{GS} and App.~A of Ref.~\cite{GSlarge-m} that
the potentials can be reorganized such that, keeping the constant spurion $\sigma_0$,
\be
\label{eq:Vexp}
V_i(\tau-\sigma_0)&=&\sum_{n=0}^\infty c_{i,n}(\tau-\sigma_0)^n\ ,\\
c_{i,n}&\sim&(n_f-n_f^*)^n\sim\delta^n\ ,\nn
\ee
introducing an infinite number of LECs $c_{i,n}$.\footnote{Each of these LECs can be
expanded in powers of $(n_f-n_f^*)$, with the expansion of $c_{i,n}$ starting at
$(n_f-n_f^*)^n$ \cite{GS}.   However, at fixed $n_f$, no information about the coefficients in
this expansion is available.}
For $V_d$, this form depends on some mild assumptions \cite{GSlarge-m}.
The key point here is that the $n$-th term in the expansion
is of order $\delta^n$. This provides a power counting such that at any given order only a finite
number of the $c_{i,n}$ come into play, thereby making the EFT predictive.
With $\sigma_0$ transforming as in Eq.~(\ref{eq:sigmatrans}), the action is invariant under
dilatations, treating the $c_{i,n}$ as additional invariant LECs.   Setting $\sigma_0=0$ recovers the
action of  Ref.~\cite{GS}.\footnote{The dilatation transformation of $\sigma_0$
reproduces the spurion transformation rules of the coefficients $c_{i,n}$ discussed in
App.~C of Ref.~\cite{GS}.}
With the power counting of Eq.~(\ref{ep-delta-counting}), to lowest order we need to keep only the terms with $n=0$ in the potentials
$V_{\pi,\tau,m}$, and, moreover, we can choose $c_{i,0}=1$ for these three potentials,
as the overall normalization can be absorbed into the LECs $f_\pi$, $f_\tau$ and $B_\pi$.  Terms
with $n$ larger than zero contribute at higher orders in the expansion in the small
parameter $\delta$.

The potential $V_d$ needs to be considered more closely.   The leading-order dilaton potential for zero quark mass is given by
\be
\label{eq:Vcl}
V_{\text{cl}}(\tau-\sigma_0)&=&f_\tau^2 B_\tau  e^{4\sigma_0}e^{4(\tau-\sigma_0)}V_d(\tau-\sigma_0)\\
&=&f_\tau^2 B_\tau  e^{4\sigma_0}e^{4(\tau-\sigma_0)}\sum_{n=0}^\infty c_{d,n}(\tau-\sigma_0)^n\ ,\nn
\ee
where, as in Eq.~(\ref{eq:Vexp}),
\begin{equation}
\label{eq:cdncounting}
c_{d,n}\sim (n_f-n_f^*)^n \sim \delta^{n}\ .
\end{equation}
The exponential $e^{4(\tau-\sigma_0)}$ dominates $V_\text{cl}(\tau-\sigma_0)$ for $\tau-\sigma_0\to\pm\infty$ because of Eq.~(\ref{eq:cdncounting}), and thus, to ensure that $v\equiv\langle\tau\rangle$ is
finite, we assume that $\lim_{\tau-\sigma_0\to\infty}V_d(\tau-\sigma_0)>0$. In addition, it is assumed that the minimum of the dilaton potential is unique; at this minimum, $ V_\text{cl}(v)<0 $ \cite{GSlarge-m}.
It follows that by choosing
\be
\sigma_0=-v \ ,
\label{sigma0}
\ee
we can arrange that $\langle\tau\rangle=0$.   The upshot is that, for $m=0$, we
can set $v=0$. The choice of $\sigma_0$ defines the origin of the $\tau$ axis.
Other choices will lead to exponential factors which in turn
can be absorbed into the LEC's \cite{GS}.

With the choice (\ref{sigma0}) for $\sigma_0$ (and a rescaling of $B_\tau$),
to order $\delta\sim n_f-n_f^*$ the classical potential for the dilaton
in the chiral limit takes the form
\be
\label{eq:Vcl2}
V_{\text{cl}}=f_\tau^2 B_\tau\,e^{4\tau}\,c_1\left(\tau-\frac{1}{4}\right) \ ,
\ee
where we set $c_1\equiv c_{d,1}$, and the field $\tau$ does not have
a zero mode, because $v=\langle\tau\rangle=0$.
Introducing the dynamical pion fields $\xi(x)$ through
\begin{equation}
U(x)=U_0 \exp\bigg(i\sqrt{2}\xi(x)\bigg)\ ,
\label{Ufactor}
\end{equation}
we expand the effective Lagrangian of Eq.\ (\ref{Leff}) in terms of $\xi(x)$  and $\tau(x)$ and assess the magnitude of each term through the Gaussian damping in the path integral. This is a stepping stone towards determining the order of all terms in the effective theory. In the following, the fermion mass is chosen to obey $\epsilon$-counting and we will use $v(m=0)=0$ as found above.
Setting $\chi=mI$, the resulting Lagrangian terms up to quadratic in the fields are
\begin{align}
\Lar_\pi &= \frac{f_\pi^2}{2}\, \Tr[\partial_\mu \xi \partial^\mu \xi] \label{Lpi}\ , \\
\Lar_\tau &= \frac{f_\tau^2}{2}\,\partial_\mu \tau \partial^\mu \tau \label{Ltau}\ ,  \\
\Lar_m &= -f^2_\pi B_\pi  m\,  {\rm Re}\, \Tr[U_0(1-\xi^2)] \label{Lm}\ , \\
\Lar_d &=f^2_\tau B_\tau\,c_1\left(-\frac{1}{4}+2\tau^2\right)\ . \label{Ld}
\end{align}
Terms linear in $\xi$ and $\tau$ do not appear in these expressions
because the zero modes have been separated out explicitly, and thus $\int \dd^4 x \ \xi(x)=\int\dd^4 x\ \tau(x)=0$.

From Eq.~(\ref{Ld}), the mass of the dilaton in the chiral limit is found to be
\begin{equation}
M_\tau^2=4c_1 B_\tau\ .
\label{eq:Mtau}
\end{equation}
The fluctuations in the dilaton field are limited by both the dilaton kinetic term ${\cal L}_\tau$ and the dilaton potential $ \Lar_d $ in the partition function to
\begin{equation}
\tau\sim\epsilon\ .
\end{equation}
Likewise the kinetic term $ \Lar_\pi $ limits the fluctuations in the pion fields to
\begin{equation}
\xi\sim\epsilon\ .
\end{equation}
We note that the $\xi$-dependent part of ${\cal L}_m$ is of order $\epsilon^6$, while
all other terms shown in Eqs.~(\ref{Lpi}-\ref{Ld}) are of order $\epsilon^4$, not counting
the constant term in ${\cal L}_d$.

\begin{boldmath}
\section{The $\epsilon$-regime of \dchpt}
\label{sec:epsilon}
\end{boldmath}

A main result at leading order in the $\epsilon$-regime is the
possibility to evaluate the partition function explicitly. This remains
possible in the $\epsilon$-counting for \dchpt\ we are
considering where the pions are in the $\epsilon$-regime, while
the dilaton is kept in the $p$-regime.

For any physical quantity accessible to our EFT, there are two types of contributions
in terms of powers of $\epsilon$: those that depend on the fermion mass $m$, and
those that do not.   In this paper, we will be focused on those that
are dependent on the fermion mass $m$.   The motivation for this is the application
to lattice simulations with fixed $n_f$ and volume, but with varying fermion mass.

\vspace{3ex}
\subsection{Factorization of the leading order partition function}

From the discussion in the previous sections we observe that the
leading-order terms in the effective Lagrangian, Eqs.~(\ref{Lpi}-\ref{Ld}),
do not couple $U_0$ to the dynamical fields $\xi(x)$ and $\tau(x)$.
Since the measure for $U=U_0\exp(i\sqrt{2}\xi)$ factorizes \cite{GLthermo},
the leading-order partition function becomes\footnote{%
As usual, the partition function for a fixed topological sector
is defined by integrating over ${\rm U}(N_f)$ instead of $\SU(N_f)$
while inserting ${\det}^\nu(U_0)$ into the integral in \Eq{ZCPT}.
}
\be
\label{ZdCPT}
 Z_{N_f}^{(\mathdchpt)}(m) &=& e^{-{\cal V} V_{\rm cl}}\, Z_{N_f}^{(\mathchpt)}(m) \ ,
\\
\label{ZCPT}
 Z_{N_f}^{(\mathchpt)}(m) &=& \rule{0ex}{4ex}
 \int_{{\rm SU}(N_f)} DU_0 \, e^{\frac{1}{2}\,{\cal V}f_\pi^2 B_\pi\,
 e^{yv}\,\Tr\left[mU_0+m^*U_0^\dagger \right]}
 \ \equiv\ e^{-{\cal V}V_{\rm eff}} \ .
\ee
Unless otherwise stated we will take $m=m^*$ real,
but we wrote \Eq{ZCPT} such that it is valid for complex $m$ as well.
We have restored the dependence on the dilaton vacuum $v=v(m)$,
which is now determined by minimizing the potential
\be
\label{Vtot}
V = V_{\rm cl} + V_{\rm eff} \ ,
\ee
where $V_{\rm eff}$ was defined above,
and $V_{\rm cl}$ is given by \Eq{eq:Vcl2}.

We will now use this partition function to address how a small quark mass
affects the dilaton mass $M_\tau$, and how the topological susceptibility
and the chiral condensate are affected by the presence of the dilatonic meson.

\begin{boldmath}
\subsection{Effective potential and $\tau$-mass correction due to a small $m$}
\label{sec:masscorr}
\end{boldmath}

Consider the expression in the exponent of the integrand in \Eq{ZCPT},
\be
\label{eq:x}
x(v)\equiv m\mathcal{V}f_\pi^2 B_\pi\, e^{yv}\ .
\ee
In the $\epsilon$-regime $x$ is $\mathcal{O}(1)$.
Therefore, $V_{\text{eff}}$ defined by \Eq{ZCPT} takes on
the form $-h(x)/{\cal V}$, where $h$ is an $\mathcal{O}(1)$ function of $x$.%
\footnote{For general $N_f$ no explicit expression for $h(x)$ is known,
to the best of our knowledge.}
The potential~(\ref{Vtot}) then takes the form
\begin{equation}
\label{eq:Valt}
V(v)=f_\tau^2 B_\tau c_1\,e^{4v}\left(v-\frac{1}{4}\right)-\frac{1}{\cal V}\,h(x(v))\ ,
\end{equation}
and $v$ solves the equation
\begin{equation}
\label{eq:saddlealt}
0=v\,e^{4v}-\frac{y}{4}\frac{x(v)}{f_\tau^2 B_\tau c_1{\cal V}}\,h'(x(v))\ .
\end{equation}
Since $f_\tau^2 B_\tau c_1{\cal V}$ is of order $1/\epsilon^2$,
it follows that the solution of this equation is of order $\epsilon^2$.
This implies that we can expand in terms of $v$, and, to order $\epsilon^2$,
we find
\begin{equation}
\label{eq:vsol}
v(m) = \frac{y}{4}\frac{x(0)}{f_\tau^2 B_\tau c_1{\cal V}}\,h'(x(0))
=\frac{y}{4}\frac{f_\pi^2 B_\pi m}{f_\tau^2 B_\tau c_1}\,h'\!\left(f_\pi^2 B_\pi m\mathcal{V}\right)\ .
\end{equation}
Later in this section we will compute the corrections to the
topological susceptibility and the chiral condensate due to $v$.
Since $v$ is of order $\epsilon^2$ these corrections enter at the NLO level
in the $\epsilon$-regime. First, however,
we turn to the effect on the dilaton mass.

The dilaton mass may be expressed in terms of the function $h$ and its derivatives:
\begin{eqnarray}
\label{eq:dilatonmassh}
M_\tau^2=\frac{V''(v)}{f_\tau^2 e^{2v}}\bigg|_{v=v(m)}
&=&4 B_\tau c_1\,e^{2v}(1+(4-y)v)
-\frac{y^2m^2 f_\pi^4 B_\pi^2 {\cal V}e^{2(y-1)v}}{f_\tau^2}\,h''(x(v))\\
&=&4 B_\tau c_1+\frac{yf_\pi^2 B_\pi m}{f_\tau^2} \left((6-y)
\,h'\!\left(f_\pi^2 B_\pi m\mathcal{V}\right)
-yf_\pi^2 B_\pi m{\cal V}\,h''(f_\pi^2 B_\pi m\mathcal{V})\right)\ ,\nn
\end{eqnarray}
where in the first line we used Eq.~(\ref{eq:saddlealt}) and in the second line
we expanded in $v\sim\epsilon^2$ and used Eq.~(\ref{eq:vsol}).
The $m$-dependent part of $M_\tau^2$ is of order $\epsilon^4$, {\it i.e.},
of order $\epsilon^2$ relative to its value in the chiral limit.

If $x\ll 1$ we can express the resulting integral in terms of a power series
in $x$ \cite{GLthermo}, and we find, for $N_f>2$ and to first non-trivial order, that
\begin{equation}
h(x)=\frac{1}{4}\,x^2\ ,
\end{equation}
yielding the effective dilaton potential for non-zero $m$,
\begin{equation}
\label{eq:V}
V(v)=f_\tau^2 B_\tau c_1\,e^{4v}\left(v-\frac{1}{4}\right)-\frac{1}{4}\mathcal{V}f_\pi^4B_\pi^2 m^2\,e^{2yv}\ .
\end{equation}
from which
\begin{equation}
v(m)=\frac{yf_\pi^4 B_\pi^2 m^2\mathcal{V} }{8f_\tau^2 B_\tau c_1}\ ,
\label{eq:vsolxsmall}
\end{equation}
which is of order $\epsilon^2x^2$.
The dilaton mass for non-zero $m$ is given by
\begin{eqnarray}
\label{eq:dilatonmass}
M_\tau^2
&=&4B_\tau c_1+(6-2y)\,\frac{yf_\pi^4 B_\pi^2 m^2\mathcal{V}}{2f_\tau^2}
\\
&=&4B_\tau c_1\left(1+(6-2y)v(m)\right)\ .\nn
\end{eqnarray}
As already mentioned,
there are also $\BigO(\epsilon^2)$ corrections to these results
that do not depend on the fermion mass $m$.   For example, one may calculate
the NLO correction to the potential $V$ of Eq.~(\ref{eq:Valt}) in the
chiral limit, and this will lead to $m$-independent NLO corrections to $v$.
However, such an NLO correction can be absorbed into a field redefinition,
by shifting $\sigma_0$ in Eq.~(\ref{sigma0}) such that $v=0$ to NLO.
In general, $m$-independent corrections are proportional to $n_f-n_f^*$, hence
the interesting NLO corrections are those that do depend on $m$, because
they can be probed by varying $m$ in numerical simulations
of a given theory, {\it i.e.}, for a fixed $n_f$.   We will thus restrict
ourselves also in the following to $m$-dependent NLO corrections only.

\vspace{3ex}
\subsection{Topological susceptibility}

As can be seen from \Eqs{ZdCPT} and~(\ref{ZCPT}),
apart from the factor of $e^{-{\cal V} V_{\rm cl}}$, the dilatonic vacuum modifies
the leading-order $\epsilon$-regime partition function by a scaling
of the mass by $e^{yv}$.  Let us promote the mass parameter to
a complex variable, $me^{i\theta}$, so that the dependence of the leading order
partition function~(\ref{ZCPT}) on $m$, $\theta$ and $v$ is now
through the combination $me^{i\theta}e^{yv}$.
The topological susceptibility\footnote{
We trust that the use of the conventional symbol $\chi$ to denote
the topological susceptibility will not lead to confusion even though
the mass spurion is also conventionally denoted by $\chi$.}
\be
\label{defchi}
\chi\equiv\frac{\langle \nu^2\rangle}{\cal V}
= -\frac{1}{{\cal V} N_f^2}
\frac{\partial^2}{\partial \theta^2}\log Z\big|_{\theta=0} \ ,
\ee
may then be obtained from the expressions for
$\langle \nu^2\rangle(m)$ derived in Ref.~\cite{LS}
by replacing $m$ with $me^{yv}$. Arriving at this result is in fact
not entirely as trivial as it would seem at first, and we refer to
App.~\ref{topodetail} for more details.

For $N_f=2$ it is possible to evaluate the partition function
and the topological susceptibility in closed form \cite{LS}. 
For higher $N_f$ (e.g.,~$N_f=8$, relevant for the near-conformal QCD-like
theories) we will content ourselves with the asymptotic form valid for
$x\gg N_f,\nu$.  The details may be found in App.~\ref{topodetail}.
The result \cite{LS,ML}
\be
\label{distnu}
\frac{Z^{(\nu,\mathdchpt)}}{Z^{(\nu=0,\mathdchpt)}} =
e^{-\frac{\nu^2}{2\langle\nu^2\rangle}} \ ,
\label{Znu}
\ee
from which, using Eq.~(\ref{Znufinal}), we obtain the asymptotic result
\be
\chi = m f_\pi^2 B_\pi e^{yv(m)}/N_f\ .
\label{nuSQ}
\ee

\subsection{Chiral condensate \label{sec:cond}}

Similarly we may consider the effect of including the dilaton on the chiral condensate
\be
\label{cond}
\langle \bar{\psi}\psi\rangle^{(\mathdchpt)}(m)
& = & \frac{1}{\cal V} \frac{\partial}{\partial m}\log Z^{(\mathdchpt)} \ ,
\ee
in the $\epsilon$-regime.
Note that $Z^{(\mathdchpt)}$ depends on $m$ also through $v(m)$.
The terms resulting from the $m$-dependence of $v$, however, take the form
\be
\frac{\partial v}{\partial m} \frac{\partial}{\partial v} \log Z^{(\mathdchpt)}
=-{\cal V} \frac{\partial v}{\partial m} \frac{\partial V}{\partial v}  \ ,
\ee
where $V$ is given by \Eq{Vtot}.
This vanishes, because $\partial V/\partial v=0$ 
at the minimum of the potential.
Therefore, as for the topological susceptibility,
the impact of the dilaton enters through
the scaling of the mass $m\to me^{yv}$ in \Eq{ZCPT},
and the chiral condensate is found to be
\be
\label{psibarpsi}
\langle \bar\psi\psi\rangle^{(\mathdchpt)}(m)
=  e^{yv}\langle \bar\psi\psi\rangle^{(\mathchpt)}(me^{yv}) \ .
\ee
Here $\langle \bar{\psi}\psi \rangle^{(\chi\rm{PT})}(m)$ is the result
from standard \chpt\ in the $\epsilon$-regime, see Refs.~\cite{GLthermo,LS}.
Once again, $v=v(m)$ is the minimum of the potential~(\ref{Vtot}).

The scaling relation~(\ref{psibarpsi})
is in fact valid in complete generality for any mass $m$
(in the domain of validity of \dchpt),
as long as the dilaton is in the $p$-regime.
When also the pions are in the $p$-regime
we arrive at the leading order result
\be
\label{pregimecond}
\langle \bar\psi\psi\rangle^{(\mathdchpt)}(m)
=  e^{yv} N_f f_\pi^2 B_\pi \ .
\ee
For pions in the $\epsilon$-regime we may expand
$e^{yv}\approx 1+yv=1+\BigO(\epsilon^2)$, and
to leading order we recover the universal result of Refs.~\cite{GLthermo,LS}.
The $\BigO(\epsilon^2)$ term coming from the expansion of $e^{yv}$
adds up to the other NLO corrections of ordinary \chpt\ \cite{DDF,ABL,AP,Damgaard:2008zs}.

Note that because there are $N_c N_f$ fermionic degrees of freedom,
the condensate scales like $N^2$ in the Veneziano limit.
Remembering that $f_\pi^2$ scales like $N$, the correct scaling
is manifest in \Eq{pregimecond} for $p$-regime pions.
For a \chpt\ derivation when the pions are in the $\epsilon$-regime,
see App.~\ref{largeN}.

\section{The Dirac spectrum from \dchpt \label{sec:Dirac}}

In this section we derive the microscopic limit of the spectral density
of the massless Dirac operator, for
fixed topological charge $\nu$ in \dchpt, by connecting it to the results
of Ref.~\cite{DOTV} for standard \chpt.
Before considering \dchpt\ we briefly review the general setup for computing
the eigenvalue density of the massless, anti-hermitian Dirac operator.

The spectral density of the Dirac eigenvalues $\Lambda_k$ is given by
\be
\label{rho}
\rho(\Lambda,m) = \sum_k \left\langle\delta(\Lambda-\Lambda_k)\right\rangle \ ,
\ee
where $m$ in the argument of $\rho$ refers to the mass of the physical
({\it i.e.}, sea) quarks  entering the average.
To compute the spectral density we first note that it can be written
as the discontinuity across the imaginary axis in the $m'$ plane,
see {\it e.g.}~Ref.~\cite{DOTV},
\be
\label{disc}
\text{Disc}|_{m'=-i\Lambda}\Sigma_{PQ}(m',m)
=\lim_{\kappa\to0}\Sigma_{PQ}(-i\Lambda+\kappa,m)-\Sigma_{PQ}(-i\Lambda-\kappa,m)
=2\pi\rho(\Lambda,m) \ ,
\label{eq:specden}
\ee
where the partially quenched condensate is defined as
\be\label{eq:pqcondensate}
\Sigma_{PQ}(m',m) = \left\langle \sum_k \frac{1}{i\Lambda_k+m'} \right\rangle \ ,
\ee
with $m'$ a new parameter independent of $m$.
The partially quenched condensate can be computed from
the graded generating functional $Z^{(\nu)}_{N_f+1|1}(m,m'|m'')$,
\be
\Sigma_{PQ}^{(\nu)}(m',m)
= \frac{1}{\cal V} \lim_{m''\to m'}\frac{\partial}{\partial m'}
\log Z^{(\nu)}_{N_f+1|1} (m,m'|m'') \ ,
\label{defSigmapq}
\ee
where $m'$ and $m''$ are the masses of respectively the valence fermionic
and bosonic (or ``ghost'') quarks (for an introduction to the graded method in chiral perturbation theory, see Ref.~\cite{GLH}).
The advantage of expressing the eigenvalue density in terms of the
graded generating functional is that the latter can be evaluated directly
in the effective theory.

The microscopic limit of the spectral density is defined as
\be
\rho_s(u,\mu)=\lim_{\cal V\to \infty} \frac{1}{{\cal V}\Sigma}\,\rho\left(\frac{u}{{\cal V}\Sigma},\frac{\mu}{{\cal V}\Sigma}\right) \ ,
\label{rhodef}
\ee
where $u=\Lambda {\cal V} \Sigma$ and $\mu=m{\cal V}\Sigma$, with $\Sigma=f_\pi^2 B_\pi$.
This limit corresponds to the thermodynamic limit in
the $\epsilon$-regime, hence if $Z^{(\nu)}_{N_f+1|1}(m,m'|m'')$ is evaluated
in the $\epsilon$-regime it is automatically $\rho_s$ that is computed.
In more detail,
since the smallest eigenvalues of the Dirac operator are of order $1/{\cal V}$
when chiral symmetry is broken \cite{BC}, and the valence-ghost sector
is introduced to probe the smallest eigenvalues, we will always choose the
valence and ghost quark masses $m'$ and $m''$ in the generating functional
to be of order $1/{\cal V}$.  Therefore, also the associated pions are
by construction in the $\epsilon$-regime.
It is important to notice
that this requirement puts no restriction on the physical (``sea'')
quark mass $m$, and the sea pions can be either in the $\epsilon$-regime
or in the $p$-regime.

We will now show that the partially quenched partition function
obeys a scaling relation generalizing \Eq{psibarpsi}, namely,\footnote{%
  To avoid cumbersome notation we omit the superscript $\nu$.
}
\begin{equation}
\label{pqcond}
\Sigma^{(\mathdchpt)}_{PQ}(m',m)
= e^{yv}\,\Sigma^{(\mathchpt)}_{PQ}(e^{yv}m',e^{yv}m)\ ,
\end{equation}
and a similar scaling relation follows for the spectral density
\begin{equation}
\label{yrho}
\rho^{(\mathdchpt)}(\Lambda,m)
= e^{yv}\,\rho^{(\mathchpt)}(e^{yv}\Lambda,e^{yv}m)\ .
\end{equation}
In both \Eq{pqcond} and \Eq{yrho}, $v$ is the minimum of the
potential~(\ref{Vtot}) in the unquenched case.
A first hint comes from the standard relation between the spectral density
and the condensate,
\be
\label{rhocond}
  \langle \bar\psi\psi\rangle(m)
  = \int d\Lambda\, \frac{\rho(\Lambda,m)}{i\Lambda+m} \ .
\ee
This relation holds both in \chpt\ and in \dchpt.
Since we already know that the \dchpt\ condensate obeys
the scaling~(\ref{psibarpsi}), it is hard to see how this will always be
compatible with \Eq{rhocond}, {\em unless} the spectral density
scales according to \Eq{yrho}.

In order to prove \Eq{yrho} we first notice that the classical solution,
$v=v(m,m',m'')$, is now determined by minimizing the potential~(\ref{Vtot})
for each $m,m',m''$, where again $V_{\rm eff}$ is defined by \Eq{ZCPT},
except with $Z_{N_f+1|1}^{(\mathchpt)}(m,m'|m'')$ replacing
$Z_{N_f}^{(\mathchpt)}(m)$.
When we apply $\partial/\partial m'$ in \Eq{defSigmapq} we again encounter
a term proportional to $\partial V/\partial v$, which, once again,
vanishes at the minimum of the potential,
and \Eq{pqcond} follows.

Next, notice that when we now set $m''=m'$ the valence and ghost sectors
cancel each other in the partition function,
and the classical solution, $v$, reduces to that of
the unquenched case, as determined by \Eq{eq:saddlealt}.
Because $v$ no longer depends on $m'$,
when we analytically continue in $m'$ and finally calculate
the discontinuity~(\ref{disc}), $e^{yv}$ plays the role
of an innocuous, real multiplicative factor,  and \Eq{yrho} readily follows.

Our reasoning leading to \Eq{pqcond} was completely general.
As an explicit check of the transition from \Eq{pqcond} to \Eq{yrho},
we consider the limit $x=m{\cal V}f_\pi^2B_\pi e^{yv}\gg 1$.
In this limit, the $\epsilon$-regime overlaps with the $p$-regime
\cite{GLthermo}, and all pion masses containing at least one sea quark
will be large compared to pion masses containing only valence and ghost quarks,
allowing us to treat those fluctuations as if they are in the $p$-regime.
This limit is interesting, since present-day simulations
of confining near-conformal theories are so far limited to sea pions
in the $p$-regime \cite{Aoki:2016wnc,Appelquist:2016viq,Appelquist:2018yqe,Fodor:2017nlp,Fodor:2019vmw}.

Assuming $x\gg1$ (as well as $N_f\gg 1$), we show in App.~\ref{largeN} that
the partially-quenched generating functional of ordinary \chpt\ factorizes,
\be
\label{factorize}
  Z^{(\nu)}_{N_f+1|1}(x,x'|x'')=Z^{(\nu)}_{N_f}(x)Z^{(\nu)}_{1|1}(x'|x'') \ .
\ee
Here $x'=m'{\cal V}f_\pi^2 B_\pi e^{yv}$ and $x''=m''{\cal V}f_\pi^2 B_\pi e^{yv}$.
As a result, the dilaton potential becomes
\begin{equation}
\label{eq:Vgraded}
V(v)=f_\tau^2 B_\tau c_1\,e^{4v}\left(v-\frac{1}{4}\right)
-\frac{1}{\cal V}\,h^{(\nu)}_{N_f}(x) -\frac{1}{\cal V}\,h^{(\nu)}_{1|1}(x'|x'')\ ,
\end{equation}
where $h^{(\nu)}_{N_f}(x(v))$ replaces $h(x(v))$ in Eq.~(\ref{eq:Valt}),
and \cite{DOTV,GSS}
\begin{eqnarray}
\label{explicith11}
h^{(\nu)}_{1|1}(x'|x'')&\equiv& \log Z^{(\nu)}_{1|1}(m'|m'')  \\
&=&\log\left[x'I_{\nu+1}(x')K_\nu(x'')+x'' I_\nu(x')K_{\nu+1}(x'')\right]
\ .\nonumber
\end{eqnarray}
The minimum of the dilaton potential, $v=v(m,m',m'')$,
is now determined by the saddle-point equation
\begin{equation}
\label{eq:saddlealtPQ}
0=v\,e^{4v}-\frac{y}{4}\frac{x}{f_\tau^2 B_\tau c_1{\cal V}}\,\frac{\partial }{\partial x}h^{(\nu)}_{N_f}(x)-\frac{y}{4}\frac{1}{f_\tau^2 B_\tau c_1{\cal V}}\, \left(x'\frac{\partial}{\partial x'}+x''\frac{\partial}{\partial x''}\right)h^{(\nu)}_{1|1}(x'|x'') \ .
\end{equation}
Using \Eq{explicith11}, it follows that the last term in \Eq{eq:saddlealtPQ}
vanishes for $m'=m''$.   As expected, $v$ is now determined entirely by
the sea-quark sector, corroborating the transition from
\Eq{pqcond} to \Eq{yrho}.

\section{Conclusions}
\label{sec:conc}

In this paper we have considered chiral perturbation theory extended
to include a light dilatonic meson. This systematic low energy effective theory
for QCD-like theories was introduced by two of us \cite{GS}
and previously studied in the $p$-regime. Here we defined and studied
the $\epsilon$-regime of this effective theory.
In this new counting the pions are in the standard $\epsilon$-regime
while the dilaton is in the $p$-regime, such that
the dilaton potential is probed only close to its minimum.
We have calculated the quark mass dependence of the effective potential,
the dilaton mass, the topological susceptibility, the chiral condensate
and the average spectral density of the Dirac operator.
At leading order the universal results from ordinary chiral perturbation theory
are recovered. The effect of the dilaton enters at the next to leading order
in the $\epsilon$-counting.
In particular, we find that the \dchpt\ condensate and spectral density
are related to their \chpt\ counterparts via a simple scaling 
of the quark mass by the quark-mass dependent classical solution
of the dilaton potential. This scaling relation is valid
for any sea-pion mass, including both the $\epsilon$-regime and the $p$-regime.

The fact that, when the sea pions are in the $\epsilon$-regime,
the eigenvalue density of the Dirac operator
at leading order is given by the universal result and that the effect
of the dilaton only enters at next to leading order,
could appear to be in contrast with the power law behaviour
expected for a scale invariant theory, see e.g. Ref.~\cite{DeGrand}.
However, note that we assume that chiral symmetry is broken spontaneously
and hence, by the Banks--Casher relation, the eigenvalue density at the origin
is non-zero. Furthermore, we assume that the dilaton is
in the $p$-regime, and hence it is natural that the generating functional
for the eigenvalue density is dominated by the much lighter
sea (and valence) pions.

It would be most interesting to contrast the findings of \dchpt\
against lattice studies of QCD-like theories.
The advantage when considering the Dirac spectrum is that
the smallest eigenvalues are naturally in the $\epsilon$-regime,
provided that the eigenvalue density is non-zero, as required
through the Banks--Casher relation for spontaneous breaking of chiral symmetry.
Our scaling result for the spectral density is valid for sea pions
in the $p$-regime as well, which is where all existing lattice studies
have been carried out \cite{GSlarge-m}.
Contrasting the results from lattice studies
with the results from \dchpt\ may help clarify the nature
of the additional light mode observed on the lattice.

\vspace{4ex}
\noindent
{\bf Acknowledgments:}
We would like to thank Gernot Akemann, Poul Henrik Damgaard, Thomas Ryttov, Ben Svetitsky, and Jac Verbaarschot for discussions.
This work was supported in part by the U.S.\ Department of Energy under
grant DE-SC0013682 (SFSU) and by the Israel Science Foundation under
grant no.~491/17 (Tel Aviv).

\appendix

\section{Topological susceptibility: some details \label{topodetail}}

Assuming $x\gg N_f,\nu$, and using \Eqs{ZdCPT} and~(\ref{ZCPT}), we have
\be
\label{Ztheta}
Z_{N_f}^{(\mathdchpt)}(\theta) = e^{-{\cal V} V_{\rm cl} + x N_f \cos\theta} \ ,
\ee
where $V_{\rm cl}$ is given by \Eq{eq:Vcl2}, and $x$ is defined in \Eq{eq:x}.
The right-hand side is to be evaluated at the new minimum of the
dilaton potential, which now depends on $\theta$.
Since we are only interested in small $\theta$
we may approximate $\cos\theta=1-\theta^2/2 + \cdots$.
Writing the classical solution as
\be
\label{v0v}
v=v_0+\delta v \ ,
\ee
where $v_0=v_0(m)$ is the solution for $\theta=0$,
the correction is $\delta v \propto \theta^2$.
Finally we substitute the new classical solution back into \Eq{Ztheta},
and expand to second order in $\theta$ and to first order in $\delta v$.
The linear dependence on $\delta v$ vanishes, since it is
proportional to saddle-point equation satisfied by $v_0(m)$,
as we already saw in Sec.~\ref{sec:cond} and Sec.~\ref{sec:Dirac}.
The result is
\be
\label{Zsmalltheta}
Z_{N_f}^{(\mathdchpt)}(\theta) = Z_{N_f}^{(\mathdchpt)}(0)\, e^{-x N_f \theta^2/2} \ ,
\ee
and \Eq{nuSQ} follows.

In order to evaluate $Z^{(\nu)} \equiv Z^{(\nu,\mathdchpt)}$ we start from
\be
\label{Znuint}
  Z^{(\nu,\mathchpt)} = \int_0^{2\pi} d\theta\, e^{i\nu N_f\theta}
  \int_{{\rm SU}(N_f)} DU_0 \, e^{\frac{x}{2}\,
    \Tr\left[e^{i\theta}U_0 + e^{-i\theta}U_0^\dagger \right]} \ .
\ee
We again assume $x\gg N_f,\nu$ and perform similar steps,
keeping only the $O(\theta^2)$ part in the exponential,
except that we now also perform the $\theta$ integral, finding
\be
\label{Znuapprox}
  Z^{(\nu,\mathdchpt)} = e^{-{\cal V} V_{\rm cl} + x N_f - \nu^2 N_f/(2x)} \ .
\ee
The classical solution $v_0(m)$ for $\nu=0$ is the same  as for $\theta=0$,
and, defining $\delta v$ again as in \Eq{v0v},
we now have a correction $\delta v \propto \nu^2$.
As before, when we substitute the solution back into \Eq{Znuapprox},
there is no linear dependence on $\delta v$, hence
\be
\label{Znufinal}
  Z^{(\nu,\mathdchpt)} = Z^{(\nu=0,\mathdchpt)}\, e^{-\nu^2 N_f/(2x)} \ ,
\ee
and \Eq{distnu} follows (with $\chi$ in \Eq{nuSQ}).

\begin{boldmath}
\section{Some results for large $N_f$ and/or large $x$ \label{largeN}}
\end{boldmath}

We will make use of the following representation for the partition function
of the partially-quenched theory in ordinary \chpt\ \cite{SVtoda,FA}
\be
\label{Zdet}
  {\cal Z}_{k|n}^{(\nu)}(\{z_i\})
  &=& \frac{\det {\cal A}}{\prod_{i=1}^n \prod_{j=i+1}^n (z_j^2-z_i^2) \,
      \prod_{i=n+1}^{n+k} \prod_{j=i+1}^{n+k} (z_j^2-z_i^2)} \ ,
\\
  {\cal A}_{ij} &=& z_i^{j-1} {\cal J}_{\nu+j-1}(z_i) \ ,
  \qquad i,j=1,\ldots,n+k \ . \rule{0ex}{4ex}
\nn
\ee
Here $k$ is the number of fermions (sea and valence), and
$n$ is the number of ghosts.  We define
${\cal J}_\nu(z_i) = (-1)^\nu K_\nu(z_i)$ for $i=1,\ldots,n$,
and ${\cal J}_\nu(z_i) = I_\nu(z_i)$ for $i=n+1,\ldots,n+k$.
In order to use this representation the $z_i$ must be
mutually different for $i=1,\ldots,n$, as well as for $i=n+1,\ldots,n+k$.
Degenerate limits can be taken in the end.

We first show that $h_{N_f}^{(\nu)} = -\log {\cal Z}_{N_f|0}^{(\nu)}$
scales like $N^2$ in the Veneziano limit.
In this case we need \Eq{Zdet} for $n=0$, $k=N_f$.
The degenerate limit is $z_1 = \cdots = z_{N_f} = x$,
where $x$ is given by \Eq{eq:x}.  It follows that
the $z_i$ variables all scale like $N$
(note the presence of $f_\pi^2$ in \Eq{eq:x}).
Thus, for $\nu+j=\BigO(1)$ we may use the asymptotic expansion
of $I_{\nu+j-1}(z_i)$ for large argument, while for $\nu+j=\BigO(N)$
we may use the uniform expansion for real variable \cite{DLMF}.
Either way, we find that $\log {\cal A}_{ij}$ scales like $N$
for any $i,j$, hence $\log {\rm det} {\cal A} = {\rm tr} \log {\cal A}$
scales like $N^2$,
establishing a similar scaling for $h_{N_f}^{(\nu)}$.
Finally, \Eq{cond} implies the same scaling for the condensate.

We next prove that, for $x \sim N \gg 1$ the partially-quenched
partition function $Z^{(\nu)}_{N_f+1|1}(x,x'|x'')$
factorizes, {\em cf.} \Eq{factorize}.  Once again, to apply \Eq{Zdet}
we assume that the sea masses are non-degenerate, taking if desired
the degenerate limit in the end.  We have $z_1=x''$, $z_2=x'$, and $z_i=x_{i-2}$ for $i=3,\ldots,N_f+2$.
There are thus $k=N_f+1$ sea and valence quarks, and there is $n=1$ ghost quark.
We proceed by writing ${\cal A}$ in a block form,
\be
  {\cal A} = \left( \begin{array}{cc} A&B\\C&D \end{array} \right) \ ,
\ee
where $A$ is $2\times2$, $B$ is $2\times N_f$, $C$ is $N_f\times 2$,
and $D$ is $N_f\times N_f$.  Using that $x'',x'=\BigO(1)$
and $x_1,\cdots,x_{N_f}=\BigO(N)$ the entries of $A$ are $\BigO(1)$.
The entries of $B$ are also $\BigO(1)$ for $j=\BigO(1)$,
whereas for $j=\BigO(N)$ we may use the asymptotic expansion for large order.
For the entries of $C$ we may always use the asymptotic expansion
for large argument.  The same is true for the entries of $D$
provided that $j=\BigO(1)$, while for $j=\BigO(N)$ we use the
uniform expansion for real variable.  Now employing Schur's
determinant identity, $\det {\cal A} = \det A\, \det (D - C A^{-1} B)$,
we conclude that $D$ dominates over $C A^{-1} B$, and the determinant
factorizes, $\det {\cal A} = \det A\, \det D$.

The last step is to consider the denominator in \Eq{Zdet}.
In the case of ${\cal Z}^{(\nu)}_{N_f+1|1}$,
the first factor in the denominator trivially collapses to one.
The second factor can be written more explicitly as
\be
  \prod_{i=1}^{N_f} (x_i^2 - x'^2)
  \prod_{i=1}^{N_f} \prod_{j=i+1}^{N_f} (x_j^2-x_i^2)
  =
  \prod_{i=1}^{N_f} x_i^2
  \prod_{i=1}^{N_f} \prod_{j=i+1}^{N_f} (x_j^2-x_i^2) \ ,
\ee
where on the right-hand side we used $x_i\gg x'$.
In addition, noting that $D$ is originally the diagonal block of ${\cal A}$
with index range $i,j=3,\ldots,N_f+2$, we have
\be
\det D = \det D' \prod_{i=1}^{N_f} x_i^2 \ ,
\ee
where the new matrix $D'$ is defined by
\be
  D'_{ij} = x_i^{j-1} I_{\nu+j+1}(x_i)\ , \qquad i,j=1,\ldots,N_f \ .
\ee
Putting it together we arrive at \Eq{factorize} with
\be
  Z^{(\nu)}_{1|1}(x'|x'') &=& \det A \ ,
\\
  Z^{(\nu)}_{N_f}(x_1,\ldots,x_n) &=&
  \frac{\det D'}{\prod_{i=1}^{N_f} \prod_{j=i+1}^{N_f} (x_j^2-x_i^2)} \ ,
\ee
where the final identification follows from the independence of the leading order asymptotic behaviour of the Bessel function on the index.


\end{document}